# Structural phase diagram of LaO$_{1-x}$F$_x$BiSSe: suppression of the structural phase transition by partial F substitutions

Kazuhisa Hoshi[1], Shunsuke Sakuragi[2], Takeshi Yajima[2], Yosuke Goto[1], Akira Miura[3], Chikako Moriyoshi[4], Yoshihiro Kuroiwa[4], and Yoshikazu Mizuguchi[1*]

[1]*Department of physics, Tokyo Metropolitan University, Hachioji 192-0379, Japan*

[2]*Institute for Solid State Physics, University of Tokyo, Kashiwa 277-8581, Japan*

[3]*Faculty of Engineering, Hokkaido University, Sapporo 060-8628, Japan*

[4]*Graduate School of Advanced Science and Engineering, Hiroshima University, Higashihiroshima, 739-8526, Japan*

Corresponding author: Y. Mizuguchi (mizugu@tmu.ac.jp)

Highlights

- Structural phase diagram of LaO$_{1-x}$F$_x$BiSSe was established.
- Synchrotron X-ray diffraction experiments were used to investigate the temperature evolutions of lattice constants of LaO$_{1-x}$F$_x$BiSSe.
- Low-temperature X-ray diffraction confirmed that the tetragonal structural type is maintained down to 4 K for LaO$_{0.5}$F$_{0.5}$BiSSe, in which possible nematic superconductivity has been observed.

Abstract

We have investigated low-temperature crystal structure of BiCh$_2$-based compounds LaO$_{1-x}$F$_x$BiSSe ($x$ = 0, 0.01, 0.02, 0.03, and 0.5), in which anomalous two-fold-symmetric in-plane anisotropy of superconducting states has been observed for $x$ = 0.5. From synchrotron X-ray diffraction experiments, a structural transition from tetragonal to monoclinic was observed for $x$ = 0 and 0.01 at 340 and 240 K, respectively. For $x$ = 0.03, a structural transition and broadening of the diffraction peak were not observed down to 100 K. These facts suggest that the structural transition could be suppressed by 3% F substitution in LaO$_{1-x}$F$_x$BiSSe. Furthermore, the crystal structure for $x$ = 0.5 at 4 K was examined by low-temperature (laboratory) X-ray diffraction, which confirmed that the tetragonal structure is maintained at 4 K for $x$ = 0.5. Our results suggest that the two-fold-symmetric in-plane anisotropy of superconducting states observed for LaO$_{0.5}$F$_{0.5}$BiSSe was not originated from structural symmetry lowering.

# 1. Introduction

BiCh$_2$-based (Ch = S, Se) superconductor was discovered in 2012 [1,2], which was followed by the development of related layered compounds [3-13]. The crystal structure of typical BiCh$_2$-based superconductor is an alternate stacking of an insulating layer and a BiCh$_2$ conducting bilayer, which is similar to those of high-temperature superconductors such as cuprates and iron-based superconductors (IBSCs) [14,15]. Particularly, REOBiCh$_2$-type (RE = Rare earth elements) compounds have been extensively studied due to its flexibility for elemental substitution of constituent elements. On the electronic characteristics, a parent phase of REOBiCh$_2$ is a semiconductor with a band gap. Electron carrier doping by a partial substitution of F for the O site has been used to induce metallicity and then superconductivity in the system [2,3]. On the pairing mechanisms of the superconductivity in the BiCh$_2$-based systems, there is still controversial; both conventional and unconventional mechanisms have been proposed from both theoretical and experimental studies [16]. For example, investigations on thermal conductivity, specific heat, and magnetic penetration depth have suggested a conventional model of superconductivity for BiCh$_2$-based superconductors [17-19]. However, angle-resolved photoemission spectroscopy (ARPES) study reported the observation of anisotropic superconducting gap, indicating unconventional superconductivity is emerging in NdO$_{0.71}$F$_{0.29}$BiS$_2$ [20]. Furthermore, the absence of isotope effects was observed in LaO$_{0.6}$F$_{0.4}$Bi(S, Se)$_2$, which was examined using $^{76}$Se and $^{80}$Se isotopes. This results also imply that unconventional pairing is essential for LaO$_{0.6}$F$_{0.4}$Bi(S,Se)$_2$ [21].

Recently, two-fold symmetry in the *ab*-plane (in-plane) anisotropy of the magnetoresistance was observed in superconducting states of LaO$_{0.5}$F$_{0.5}$BiSSe single crystals [22]. From room-temperature structural analysis using X-ray diffraction (XRD), the crystal structure was determined to be a tetragonal type (*P4/nmm*) having four-fold symmetry in the *ab*-plane [22,23]. Therefore, in the superconducting states, the in-plane anisotropy of magnetoresistance is expected to break its structural symmetry in the conducting plane. This phenomenon is quite similar to what observed in *nematic superconductors*, $A_x$Bi$_2$Se$_3$ (A = Cu, Sr, Nb) and IBSCs [24-29]. In those nematic superconductors, two-fold symmetric anisotropy of physical properties in superconducting states has also been observed, in spite of three-fold ($A_x$Bi$_2$Se$_3$) or four-fold (IBSCs) symmetry in its crystal structure. For LaOBiSSe, however, there is still possibility of the emergence of two-fold symmetric anisotropy in the superconducting states due to structural symmetry lowering, because the parent phase (F-free) LaOBiSSe undergoes a structural transition from tetragonal (high-*T* phase: *P4/nmm*) to monoclinic (low-*T* phase: *P*2$_1$/*m*) at 300–400 K [3,23]. Furthermore, BiS$_2$-based LaO$_{0.5}$F$_{0.5}$BiS$_2$ shows a pressure-induced structural transition from tetragonal to monoclinic [30]. Since the origin of those structural transitions can be linked to the activity of the Bi lone-pair electrons, which affect local structures of the conducting BiCh$_2$ layers [31], the structural instability could presence for all BiCh$_2$-based systems. However, it has been known that carrier doping via partial element substitution suppresses the structural transition and stabilizes tetragonal structure in a REOBiCh$_2$-type structure [3]. As well, theoretical analysis on the stability of the crystal structure as a function of carrier concentration suggested that the tetragonal structure is more stabilized than monoclinic one in electron-doped (F-substituted) LaOBiS$_2$ [32].

On the basis of those facts, systematic analyses on the crystal structure near the cross-over between tetragonal and monoclinic structure of LaO$_{1-x}$F$_x$BiSSe are, therefore, needed to understand the relationship between the two-fold-symmetric in-plane anisotropy of superconducting properties and possible nematic superconductivity. Here, we have studied temperature and carrier concentration dependences of crystal structure for LaO$_{1-x}$F$_x$BiSSe. A structural transition from tetragonal to monoclinic was observed at 340 K for $x = 0$ and at 240 K for $x = 0.01$. No structural transition and X-ray diffraction peak broadening were not observed for $x = 0.03$ down to 100 K. These results suggest that the structural transition is rapidly suppressed by carrier doping and disappears at $x \sim 0.03$ in LaO$_{1-x}$F$_x$BiSSe. Furthermore, we have confirmed that the tetragonal structure has been maintained at 4 K for LaO$_{0.5}$F$_{0.5}$BiSSe ($x = 0.5$), where anomalous two-fold-symmetric in-plane anisotropy of superconducting states was observed.

## 2. Material and methods

Polycrystalline samples of LaO$_{1-x}$F$_x$BiSSe ($x = 0, 0.01, 0.02, 0.03,$ and $0.5$) were prepared by solid-state-reaction method. Bi$_2$S$_3$ and Bi$_2$Se$_3$ were pre-synthesized by reacting Bi (99.999%), S (99.9999%), and Se (99.999%) grains. Powders of La$_2$O$_3$ (99.9%), La$_2$S$_3$ (99.9%), BiF$_3$ (99.9%), Bi$_2$S$_3$, and Bi$_2$Se$_3$, and Bi (99.999%) grains with nominal compositions of LaO$_{1-x}$F$_x$BiSSe were mixed, pressed into a pellet, sealed into an evacuated quartz tube, and annealed at 700˚C for 15 h. The obtained sample was mixed for homogenization, pressed into a pellet, sealed into an evacuated quartz tube, and annealed at 700˚C for 15 h. Synchrotron X-ray diffraction (SXRD) experiments were performed from 400 to 300 K for $x = 0$ and from 300 to 100 K under the temperature control system with nitrogen gas for $x = 0.01, 0.02,$ and $0.03$ at the beamline BL02B2 of SPring-8 under research proposals Nos. 2019A1114 ($\lambda = 0.496197$ Å) and 2019B1195 ($\lambda = 0.496391$ Å). For $x = 0.5$, high-resolution powder X-ray diffraction (XRD) experiments with a Cu-K$\alpha$1 radiation monochromatized by a Ge(111)-Johansson-type monochromator at 300 and 4 K were performed on SmartLab diffractometer equipped with a GM refrigerator. The typical XRD patterns of both SXRD and conventional XRD experiments are shown in supplemental data (Figs. S1 and S2). The temperature dependence of electrical resistivity was measured by four-terminal method with a DC current of 1 mA on a GM refrigerator.

## 3. Results and discussion

As reported in Refs. 3 and 13, the parent phase LaOBiSSe ($x = 0$) undergoes a structural transition from tetragonal (*P4/nmm*) to monoclinic (*P2$_1$/m*) at 300 – 400 K. In a monoclinic phase, the 200 peak splits into 200 and 020. Therefore, by scanning the 200 peak on temperature, the structural transition temperature ($T_s$) and the evolution of in-plane lattice constants ($a$ and $b$) can be investigated. Figures 1(a–d) show the temperature dependences of the tetragonal 200 peak and monoclinic 200 and 020 peaks on the SXRD patterns for $x = 0$, 0.01, 0.02, and 0.03, respectively. Commonly, the 200 peak shifts to higher angles on cooling, which is due to the compression of the lattice. As shown in Fig. 1(a), peak splitting of the 200 peak into the monoclinic 200 and 020 peaks was observed below 340 K for $x = 0$, which indicates a structural transition to the monoclinic structure at $T_s = 340$ K. Note that the peak intensity for $x = 0$ is rapidly suppressed with decreasing temperature as the temperature approached $T_s$. This is a signature of structural instability toward a structural transition (symmetry lowering) because the decrease in peak intensity during the temperature scanning corresponds to the broadening of the peak in this experimental setup, in which the sample condition was not modified and the temperature was continuously changed. A structural transition was observed for $x = 0.01$ at 240 K. A similar trend on the suppression of peak intensity was observed in Fig. 1(b). For $x = 0.02$ and 0.03, a clear structural transition was not observed down to 100 K. However, the suppression of the peak intensity was observed for $x = 0.02$. This signature implies that the sample has the structural instability, and a structural transition is expected at a temperature lower than 100 K. Notably, the peak intensity is almost constant from 300 to 100 K for $x = 0.03$, which implies that no structural transition is expected at temperatures lower than 100 K. To check the evolution of the peak broadening, the temperature evolutions of the full width half maximum (FWHM) of the 200 peaks estimated from the Gaussian fitting are also consistent with the scenario above (see Fig. S3 of Supplemental data). These results suggest that the structural transition can be completely suppressed at concentration lower than $x = 0.03$ in LaO$_{1-x}$F$_x$BiSSe.

To analyze lattice constants $a$ and $b$ from the data shown in Fig. 1, the 200 and 020 peaks were fitted by one or two Gaussian functions. Two Gaussian functions were used for $x = 0$ and 0.01, where a clear structural transition was observed. For $x = 0.02$ and 0.03, we analyzed the lattice constant with one Gaussian function. Figure 2(a) shows the temperature dependence of the lattice constants $a$ and $b$ for $x = 0$, which clearly shows a transition at $T_s = 340$ K. As shown in Fig. 2(b), the $T_s$ for $x = 0.01$ was 240 K. For $x = 0.02$ and 0.03, the lattice constant $a$ linearly changed with decreasing temperature, which implies that the tetragonal structure is dominant in this temperature regime. The trend that the structural transition from tetragonal to monoclinic is rapidly suppressed by F substitution in LaO$_{1-x}$F$_x$BiSSe is consistent with the theoretical study which proposed that the tetragonal structure is more stable than monoclinic in F-substituted LaOBiS$_2$ [32].

Figure 3 shows a structural phase diagram of LaO$_{1-x}$F$_x$BiSSe. Due to the experimental limitation, we could scan the lattice constant on temperature down to 100 K. From the evidence of the suppression of peak intensity, a structural transition at below 100 K for $x = 0.02$ is assumed. In contrast, since the peak intensity for $x = 0.03$ does not show a decrease down to 100 K, we assume that the low-temperature structure for $x = 0.03$ is tetragonal down to 0 K, which is indicated with a cross symbol in Fig. 3.

To investigate the influence of the structural transition to the transport properties, the temperature dependence of electrical resistivity was measured for $x = 0$ ($T_s = 340$ K), 0.01 ($T_s = 240$ K), 0.02, and 0.03 (no transition is expected) and plotted in Fig. 4. As reported in Ref. 13, an upturn, non-metallic behavior is observed below 90 K for $x = 0$. For $x = 0.01$, metallic-like behavior was observed while a small upturn is observed at low temperature. Notably, there is no clear anomaly at $T_S$, as indicated by an arrow, where a structural transition was detected (see Figs. 1 and 2). The absence of anomaly in the resistivity data is probably because of the small distortion of the in-plane structure ($a/b$ ratio in the low-$T$ phase) below the $T_S$. The $a/b$ ratio for x = 0.01 is 1.002, which indicates 0.2% in-plane distortion. For example, this is clearly smaller than 0.8% in-plane distortion that observed in layered compound $BaFe_2As_2$ [33], in which a clear anomaly is observed in its resistivity-temperature curve. For $x = 0.02$ and 0.03, a typical metallic behavior is observed. In addition, a small hump is observed at $T = 150–200$ K for $x = 0.03$. Since no structural anomaly was observed for $x = 0.03$ in the temperature range, we have no explanation about the anomaly at present. However, a similar anomaly has been reported for several Bi-chalcogenide layered compounds [34-36], and possible charge-density-wave ordering has been suggested for $EuFBiS_2$ [37].

From the results described above, it is reasonable to expect that the low-temperature crystal structure of $x = 0.5$, in which two-fold symmetric in-plane anisotropy of the magnetoresistance in the superconducting states was observed [22], is tetragonal with four-fold symmetry in $ab$-plane. To confirm this assumption, low-temperature laboratory XRD experiments were performed for $x = 0.5$ at 4 K. Figure 5(a) shows the 200 peaks at $T = 4$ and 300 K collected with a Cu-K$\alpha$ radiation using a pelletized sample. The peak at 4 K shifts to a higher angle because of lattice compression by cooling. Neither peak splitting nor broadening was observed for the 200 peak, indicating that the tetragonal structure is maintained at 4 K for $x = 0.5$. Figure 5(b) shows the 004 peaks collected at $T = 4$ and 300 K for $x = 0.5$. No peak broadening is observed for the 004 peak at 4 K. From the structural investigations for $LaO_{1-x}F_xBiSSe$ shown here, we suggest that the origin of the two-fold symmetric in-plane anisotropy of magnetoresistance in the superconducting states of $LaO_{0.5}F_{0.5}BiSSe$ is not structural one but another electronic one, as suggested for $Bi_2Se_3$-based nematic superconductors.

## 4. Conclusions

We have investigated low-temperature crystal structure of $BiCh_2$-based compounds $LaO_{1-x}F_xBiSSe$ ($x = 0, 0.01, 0.02, 0.03$, and 0.5). From SXRD experiments, a structural transition from tetragonal to monoclinic was observed for $x = 0$ and 0.01. For $x = 0.03$, a structural transition and broadening of the diffraction peak were not observed down to 100 K. These facts suggest that the structural transition could be suppressed by 3% F substitution in $LaO_{1-x}F_xBiSSe$. Furthermore, from XRD experiments at $T$ = 4 K, the crystal structure for $x$ = 0.5 at 4 K was determined as tetragonal. The structural phase diagram obtained in this study suggests that the origin of the two-fold symmetric in-plane anisotropy of magnetoresistance in the superconducting states of LaO$_{0.5}$F$_{0.5}$BiSSe is not structural one but another electronic one.

**Acknowledgments**

This work was partly supported by grants in Aid for Scientific Research (KAKENHI) (Grant Nos. 15H05886, 16H04493, 18KK0076, and 19K15291), JST-CREST (Grant No. JPMJCR20Q4), and the Advanced Research Program under the Human Resources Funds of Tokyo (Grant Number: H31-1).

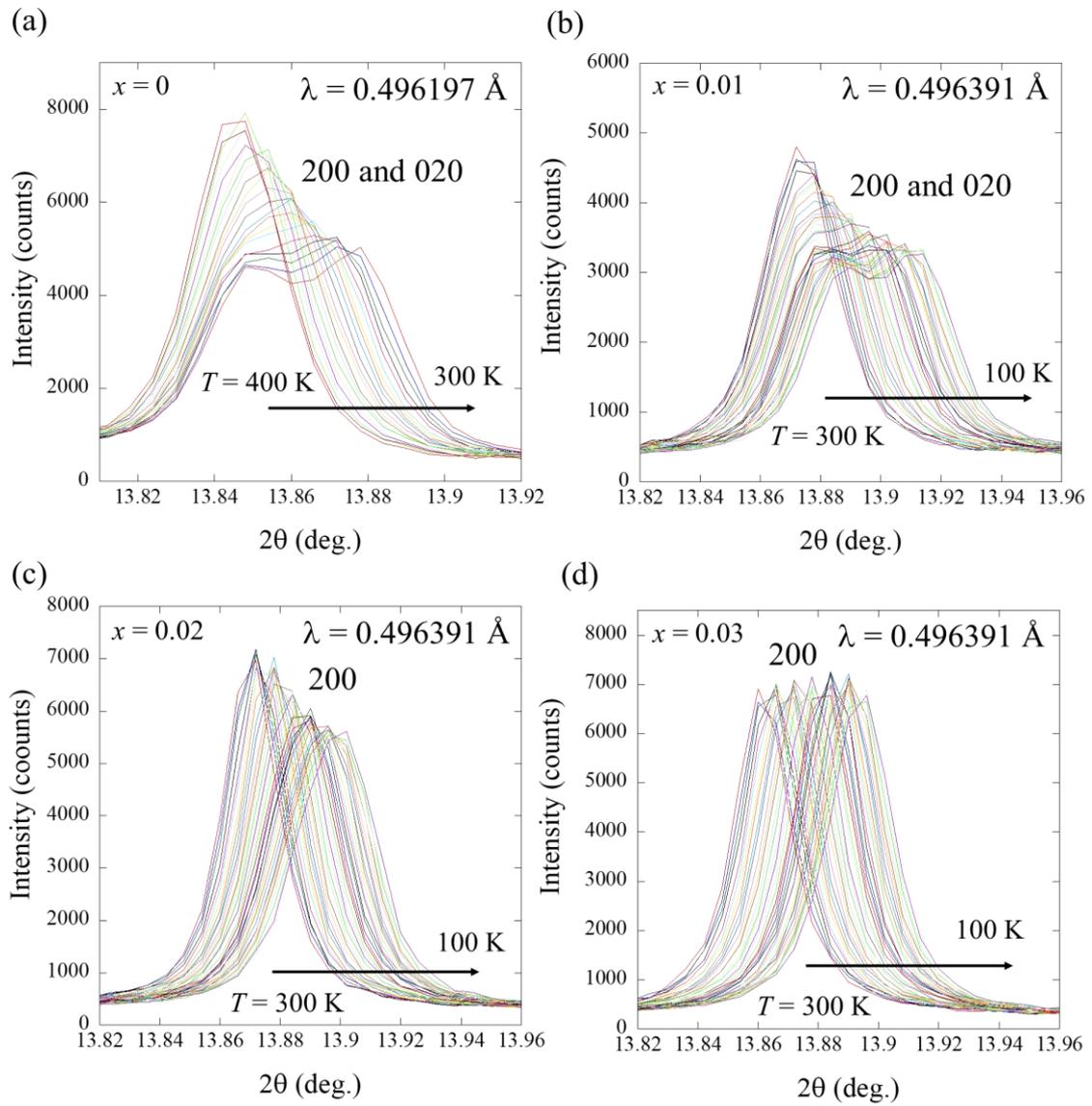

Fig. 1. Temperature evolutions of the 200 and 020 peaks of the SXRD patterns for (a) $x = 0.01$ (b) $x = 0.01$, (c) $x = 0.02$, and (d) $x = 0.03$. The wavelengths used in the scanning are indicated in the figures.

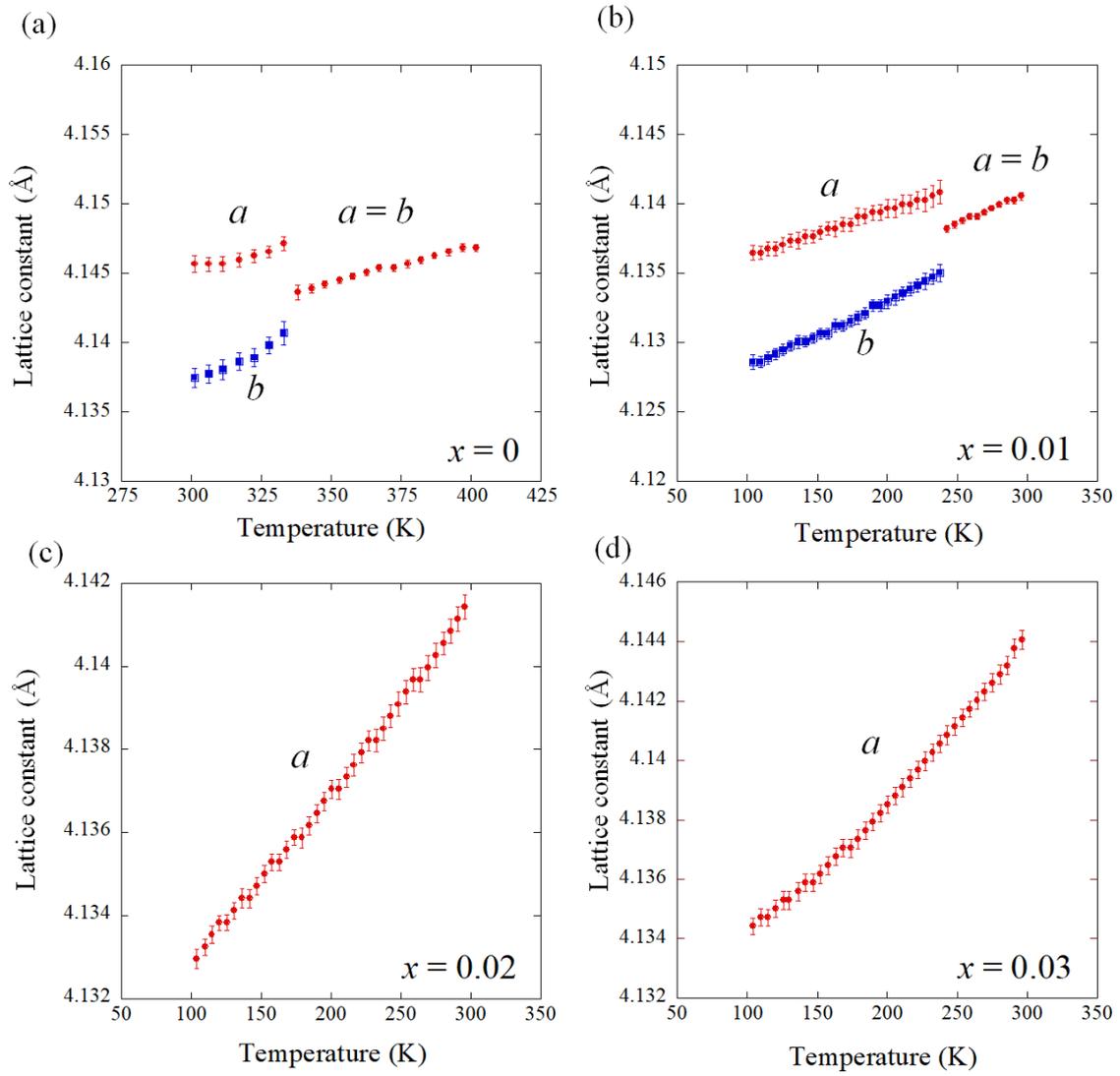

Fig. 2. Temperature dependences of the lattice constants $a$ and $b$ for LaO$_{1-x}$F$_x$BiSSe: (a) $x = 0$, (b) $x = 0.01$, (c) $x = 0.02$, and (d) $x = 0.03$.

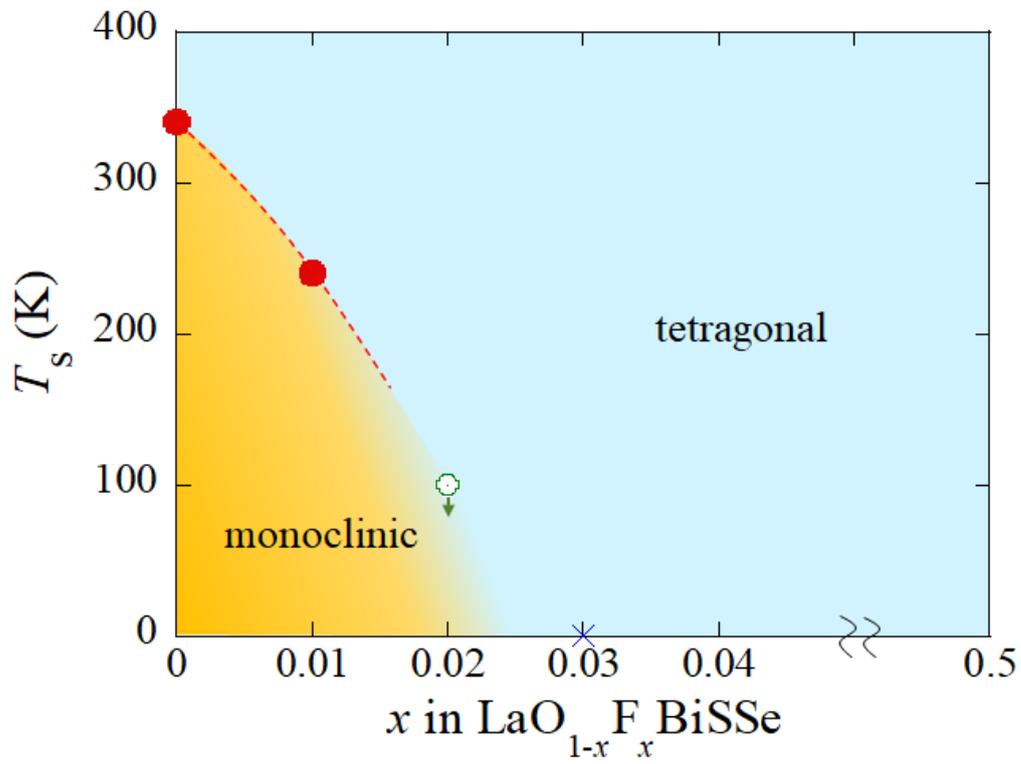

Fig. 3. Structural phase diagram of LaO$_{1-x}$F$_x$BiSSe. A circle symbol with an arrow at $x$ = 0.02 indicates that the $T_s$ for $x$ = 0.02 is lower than 100 K. A cross symbol at $x$ = 0.03 has been plotted under the assumption that the sample with $x$ = 0.03 does not undergo a structural transition down to 0 K.

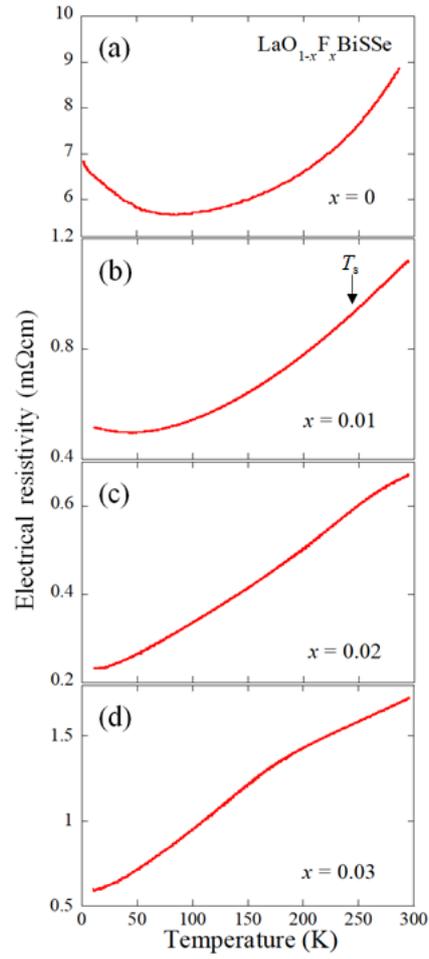

Fig 4. Temperature dependences of resistivity for (a) $x = 0$, (b) $x = 0.01$, (c) $x = 0.02$, and (d) $x = 0.03$. $T_s$ denotes the structural transition temperature for $x = 0.01$.

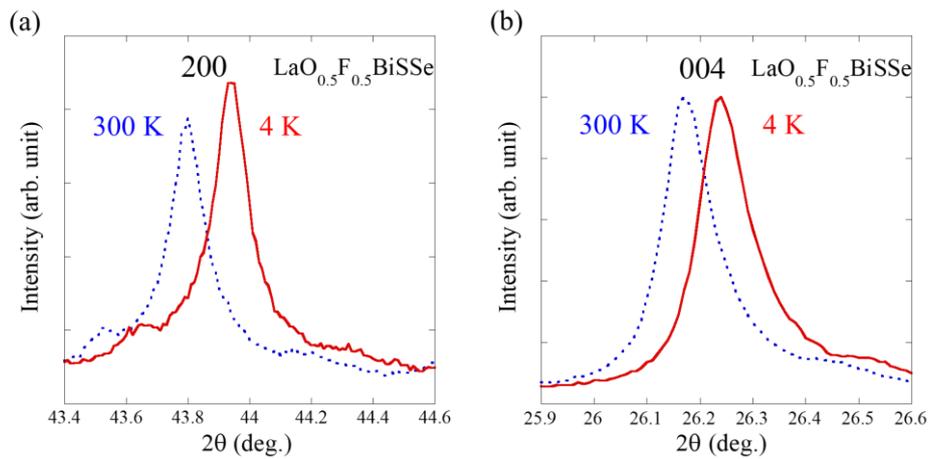

Fig 5. (a) 200 peak and (b) 400 peak in laboratory XRD pattern for LaO$_{0.5}$F$_{0.5}$BiSSe ($x = 0.5$) collected at 4 and 300 K.

# Structural phase diagram of LaO$_{1-x}$F$_x$BiSSe: suppression of the structural phase transition by partial F substitutions

Kazuhisa Hoshi[1], Shunsuke Sakuragi[2], Takeshi Yajima[2], Yosuke Goto[1], Akira Miura[3], Chikako Moriyoshi[4], Yoshihiro Kuroiwa[4], and Yoshikazu Mizuguchi[1*]

[1]*Department of physics, Tokyo Metropolitan University, Hachioji 192-0379, Japan*
[2]*Institute for Solid State Physics, University of Tokyo, Kashiwa 277-8581, Japan*
[3]*Faculty of Engineering, Hokkaido University, Sapporo 060-8628, Japan*
[4] *Graduate School of Advanced Science and Engineering, Hiroshima University, Higashihiroshima, 739-8526, Japan*

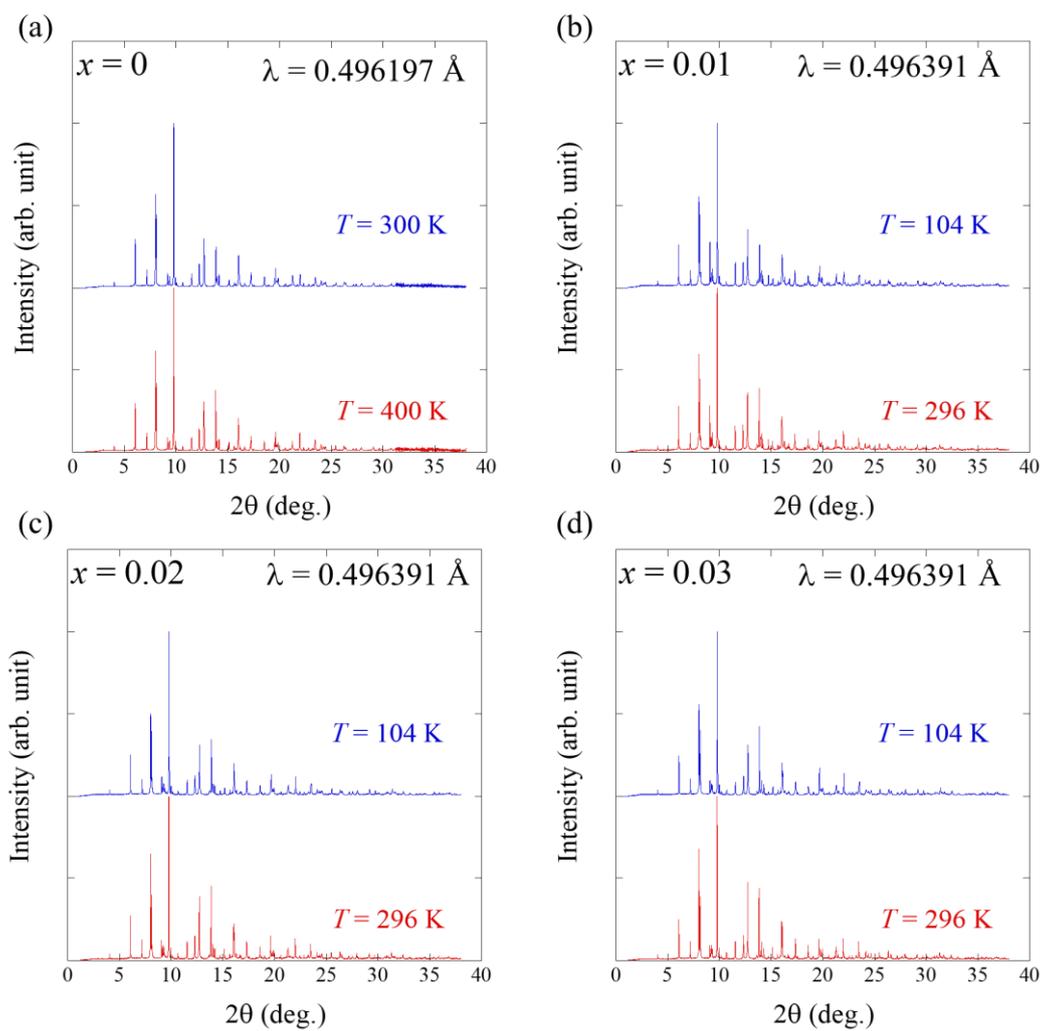

Fig. S1. Synchrotron powder X-ray diffraction (SXRD) patterns for (a) $x = 0$, (b) $x = 0.01$, (c) $x = 0.02$, and (d) $x = 0.03$ of $LaO_{1-x}F_xBiSSe$.

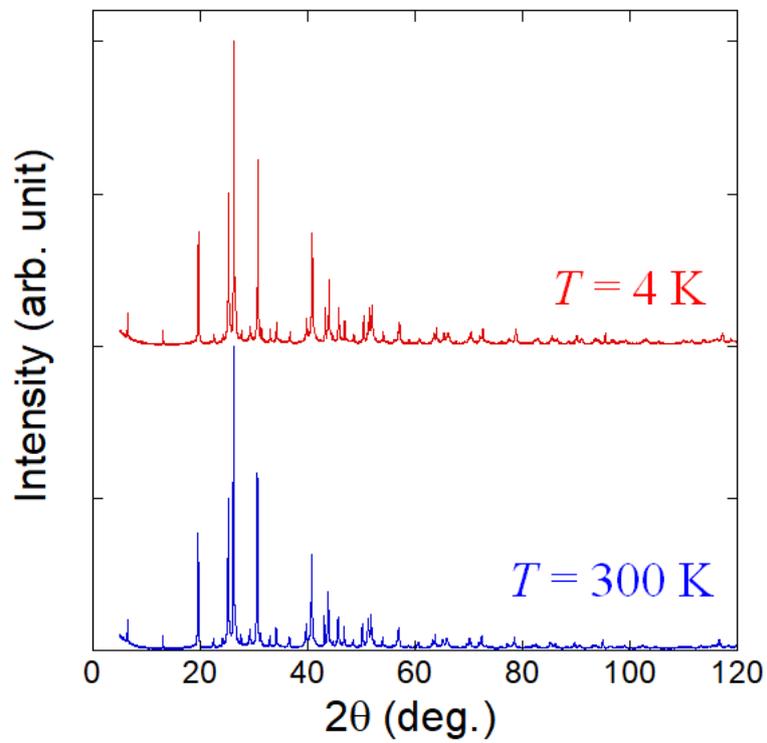

Fig. S2. Powder X-ray diffraction (SXRD) patterns for LaO$_{0.5}$F$_{0.5}$BiSSe.

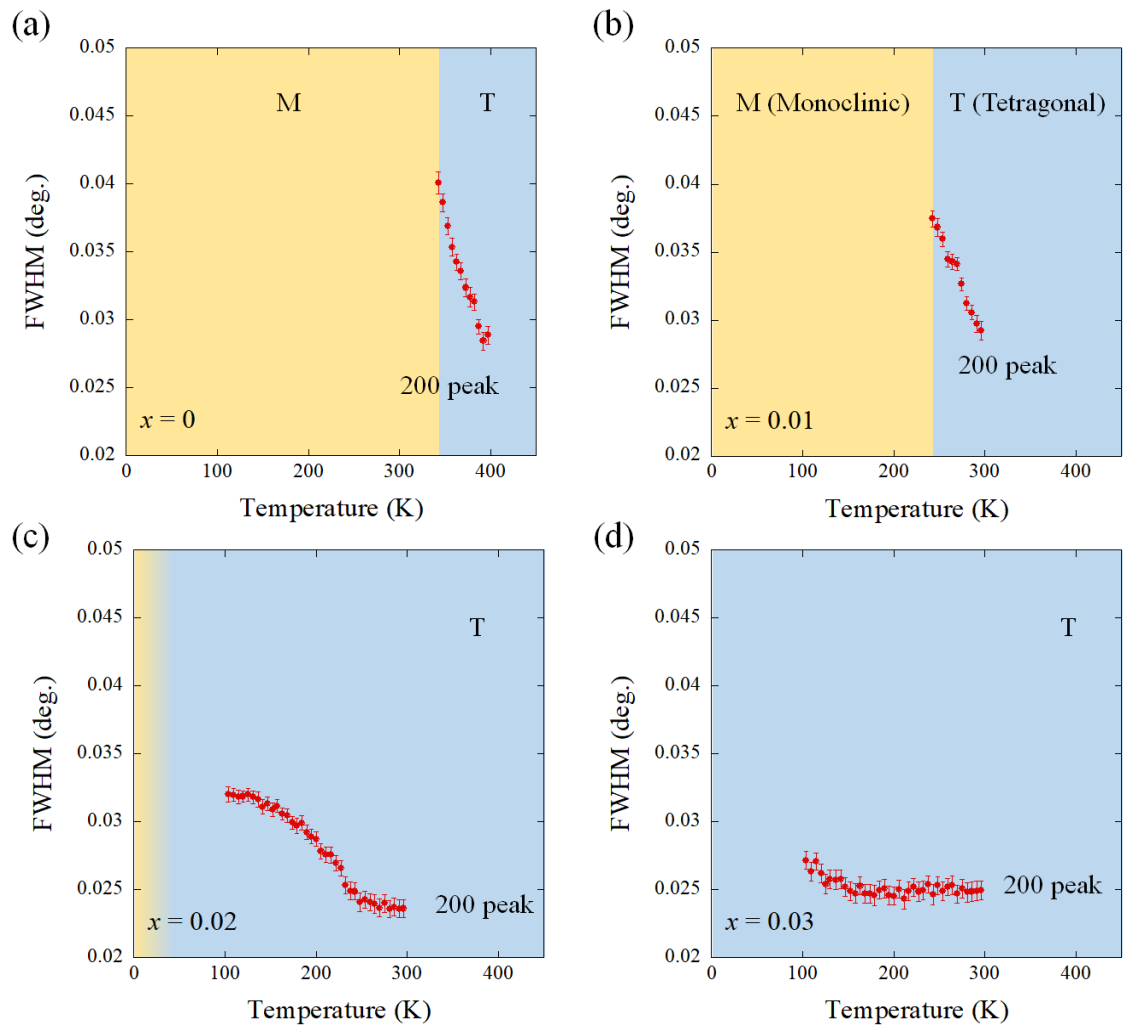

Fig. S3. Temperature evolution of full width half maximum (FWHM) of the 200 peak for (a) $x = 0$, (b) $x = 0.01$, (c) $x = 0.02$, and (d) $x = 0.03$ of LaO$_{1-x}$F$_x$BiSSe.

16